\documentclass[preprint,12pt,superscriptaddress,floatfix]{revtex4}

\usepackage{graphicx}

\begin{document}

\title{Hydrogen storage with titanium-functionalized graphene}

\author{Torge Mashoff}
\affiliation{Center for Nanotechnology Innovation @ NEST, Istituto
Italiano di Tecnologia, Piazza San Silvestro 12, 56127 Pisa Italy}

\author{Makoto Takamura}
\author{Shinichi Tanabe}
\author{Hiroki Hibino}
\affiliation{NTT Basic Research Laboratories, NTT Corporation, 3--1,
Morinosato Wakamiya, Atsugi, Kanagawa 243--0198, Japan}

\author{Fabio Beltram}
\affiliation{Center for Nanotechnology Innovation @ NEST, Istituto
Italiano di Tecnologia, Piazza San Silvestro 12, 56127 Pisa Italy}
\affiliation{NEST, Istituto Nanoscienze -- CNR and Scuola Normale
Superiore, Piazza San Silvestro 12, 56127 Pisa, Italy}

\author{Stefan Heun}
\email{stefan.heun@nano.cnr.it}
\affiliation{NEST, Istituto Nanoscienze -- CNR and Scuola Normale
Superiore, Piazza San Silvestro 12, 56127 Pisa, Italy}

\date{\today}

\begin{abstract}
We report on hydrogen adsorption and desorption on titanium--covered
graphene in order to test
theoretical proposals to use of graphene functionalized with metal atoms
for hydrogen storage. At room temperature 
titanium islands grow with an average diameter of about 10\,nm. Samples were
then loaded with hydrogen, and its desorption kinetics was studied by
thermal desorption spectroscopy. We observe the desorption of hydrogen
in the temperature range between 400\,K and 700\,K. Our results
demonstrate the stability of hydrogen binding at room temperature and
show that hydrogen desorbs at moderate temperatures in line with what
required for practical hydrogen-storage applications.
\end{abstract}

\maketitle

Hydrogen is a very attractive clean fuel since the only combustion waste
product is water.\cite{Schlapbach2001} Several technical problems hinder
its widespread use, among these its
storage.\cite{Schlapbach2001,B802882F} In this respect, graphene
recently attracted much attention as a storage medium owing to its
chemical stability, low weight, and favorable physical--chemical
properties for hydrogen adsorption.\cite{Elias30012009} The storage
capacity of graphene can be further increased by surface
functionalization.\cite{PhysRevB.87.014102} One of the most promising
candidates for such functionalization is
titanium.\cite{PhysRevB.77.085405,doi:10.1021/jp100230c} Theory predicts gravimetric densities as high as 7.8~wt\% for such systems.\cite{PhysRevB.77.085405}

Several reports exist on the behavior of metal atoms and clusters on
graphene surface both theoretical \cite{choi:153108, PhysRevB.83.205408,
PhysRevB.84.235446, PhysRevB.79.195425} and
experimental.\cite{PhysRevB.84.045431, mccreary:192101,
PhysRevB.80.075406, PhysRevLett.109.026103,
PhysRevB.85.125410,PhysRevLett.106.156101,PhysRevLett.108.156803}
Concerning its potential for hydrogen storage, a number of theoretical
results are available for various metals on graphene, such as
calcium,\cite{PhysRevB.79.075431,PhysRevB.79.041406,doi:10.1021/jp907368t,Kim2009,doi:10.1021/nl902822s,reunchan:093103,Beheshti20111561,hussain:183902}
lithium,\cite{ataca:043123,an:173101} aluminum,\cite{PhysRevB.81.205406}
or transition
metals\cite{PhysRevB.77.085405,doi:10.1021/jp100230c,PhysRevB.87.014102,Shevlin2009,Cabria2012}
and transition metal ethylene
complexes.\cite{PhysRevLett.97.226102,PhysRevLett.100.105505} 
A common feature of all these proposals is that the metal atoms (and not the graphene) store the hydrogen. Nevertheless, the role of graphene in this context is fundamental: it acts as a lightweight substrate to carry the metal atoms and contributes thereby to reach high gravimetric densities.
However, there are no experimental results
published so far which would attempt to verify these proposals. The purpose of this paper is to experimentally
investigate hydrogen-storage properties of titanium-covered graphene.

\begin{figure}
  \centerline{\includegraphics[width=0.5\columnwidth]{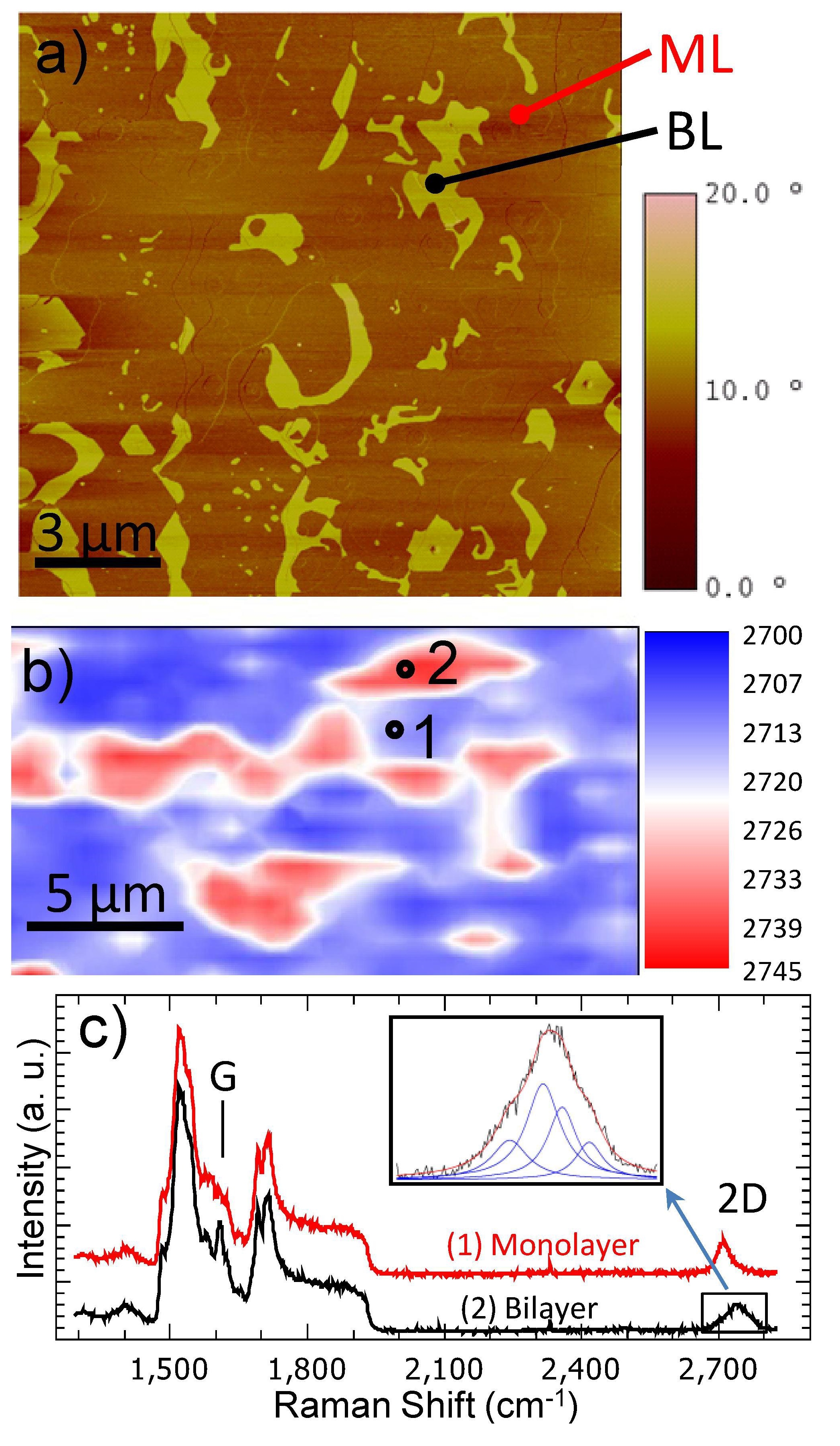}}
  \caption{(Color online) a) AFM Phase image ($14.4\times 14.4\,\mu m^2$) of one of our graphene samples, showing the phase contrast between monolayer (ML) and bilayer (BL) regions. b) Peak position map of the Raman 2D peak from one of our graphene samples. Red areas indicate regions where energy position corresponds to bilayer graphene, blue areas correspond to graphene monolayer. Based on such maps, we estimate that a fraction of 85\,\% of our samples is monolayer graphene and the remaining 15\,\% is bilayer graphene. c) Raman spectra measured at the positions indicated in (b), corresponding to monolayer (1) and bilayer (2) regions (offset for clarity). The inset shows a fit of the bilayer 2D peak with 4 Lorentzian curves.}
  \label{raman}
\end{figure}

We used monolayer graphene grown on 4H--SiC(0001) as a substrate. It was obtained by annealing 4H--SiC(0001) samples for several minutes in argon atmosphere of 100\,mbar at about 2100\,K inside a furnace. Graphene quality and the actual number of graphene layers were verified by atomic force microscopy (AFM) and Raman spectroscopy as shown in Fig.~\ref{raman}. Fig.~\ref{raman}(a) shows a $14.4 \times 14.4\,\mu m^2$ sized AFM phase image of the graphene surface of one of our samples, which shows a strong contrast between mono- and bilayer regions. Fig.~\ref{raman}(b) and (c) show Raman measurements performed on one of our samples. Raman spectra were measured under ambient conditions using an excitation wavelength of 532\,nm. The significant Raman peaks for graphene are the G peak and the 2D peak, which changes intensity and position with the number of layers. Although the exact position of the 2D peak depends in addition to the laser wavelength\cite{Ferrari2007} also on the strain in the graphene layer, which originates from the large difference in thermal expansion coefficients between graphene and SiC substrate,\cite{strudwick:051910} a shift of the 2D peak position gives a good indication for the actual number of layers. Fig.~\ref{raman}(b) shows the peak-position map for the Raman 2D band in one of our samples. Fig.~\ref{raman}(c) shows Raman spectra taken at the two positions indicated in Fig.~\ref{raman}(b). The G and the 2D peak positions are marked in the spectra. The other features originate from the SiC substrate. In our case, the position of the 2D peak is at approximately $2710\,cm^{-1}$ for monolayer graphene, and at $2740\,cm^{-1}$ for bilayer graphene. The inset to Fig.~\ref{raman}(c) shows a fit of the bilayer 2D peak with four Lorentzian curves of the same full width at half maximum (FWHM) of $35\,cm^{-1}$ which confirms the assignment of this area to bilayer graphene.\cite{Graf2007} An analysis of the different regions shows that about 85\,\% of the sample is monolayer graphene (blue area in the peak--position map) and 15\,\% is bilayer graphene (red). A similar result was obtained when analyzing Raman FWHM maps of the 2D band.

\begin{figure}
  \centerline{\includegraphics[width=0.5\columnwidth]{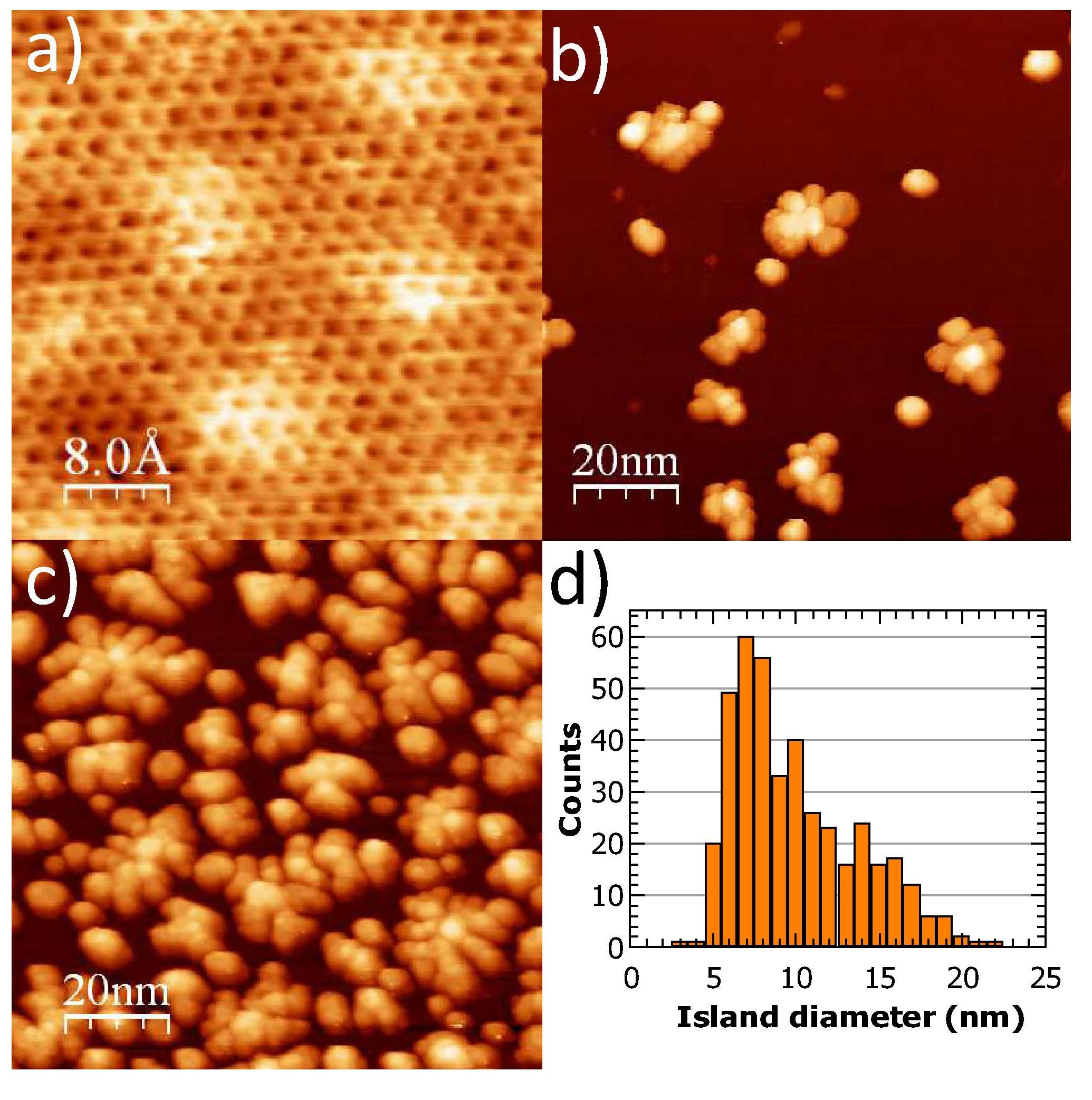}}
  \caption{(Color online) a) Hexagonal graphene lattice ($4\times 4$\,nm$^2$) on the clean surface prior to titanium deposition (V\,=\,$-0.6$\,V, I\,=\,290\,pA). b) STM image of a $100\times 100$\,nm$^2$ graphene area with a titanium coverage of 16\,\% or 0.56 ML (V\,=\,2\,V, I\,=\,280\,pA). c) $100\times 100$\,nm$^2$ graphene area with a titanium coverage of 79\,\% or 3.3 ML. d) Size distribution of the islands at 16\,\% Ti coverage (corresponding to the surface shown in (b)).}
  \label{STM_image}
\end{figure}

Hydrogenation experiments and further sample characterization were performed in an UHV chamber with a base pressure of
$5\times 10^{-11}$\,mbar, with a variable-temperature scanning tunneling
microscope (STM). The chamber is also equipped with $H_2$ supply,
Ti--evaporator, heating stage, and quadrupole mass spectrometer. Details
about the microscope are described elsewhere.\cite{Goler2013249} Before
titanium deposition, we annealed the sample at 900\,K for several hours
to remove adsorbents and to obtain a clean surface. This was done by
direct current heating to ensure a homogeneous sample temperature. The high quality of the pristine graphene films is  
testified by atomically-resolved STM images such as the one in Fig.~\ref{STM_image}(a) 
which were routinely obtained and show the hexagonal lattice of graphene.
Titanium was deposited on graphene at room temperature using a commercial
electron-beam evaporator. The Ti--coverage was calibrated by STM
imaging. All temperatures were measured using a thermocouple mounted on
the sample holder, directly in contact with the sample and additionally
cross--calibrated with a pyrometer. For a better signal--to--noise ratio
we used deuterium ($D_2$, mass 4) rather than hydrogen ($H_2$, mass 2).
Deuterium is chemically identical to hydrogen; however, we caution the
reader that the desorption temperatures might be slightly shifted owing
to the well known isotope effect.\cite{gdowski:1409,Zecho2002} During
thermal desorption spectroscopy (TDS), the mass 4--channel of the mass
spectrometer was recorded.

As shown in the STM images in Fig.~\ref{STM_image}(b) and (c), titanium atoms form
small islands on graphene and these islands grow and become more dense
at higher coverages.  Figure~\ref{STM_image}(b) shows data taken with a
surface coverage of 16\,\%, i.e.~16\,\% of the surface is covered by
Ti--islands, while the remaining 84\,\% of the surface is bare graphene.
Numerically obtaining the total volume of the Ti--islands from such images, and
assuming a layer distance in the islands as in hexagonal closed packed bulk Ti ($d=0.2342$\,nm),\cite{Wood783} this equals 0.56 monolayers (ML)
of titanium. Figure~\ref{STM_image}(c) refers to a sample with a higher Ti--coverage (79\,\%, 3.3\,ML): the average size of the islands as well as their number is increased. The histogram in Fig.~\ref{STM_image}(d) shows the distribution of the island diameters for a surface with 16\,\% Ti--coverage (cf.~Fig.~\ref{STM_image}(b)). Many islands have a diameter between 5\,nm and 9\,nm, while a few reach diameters up to 22\,nm, resulting in an average island diameter of around 10\,nm for this Ti--coverage. A careful
analysis of our STM images shows that for a small amount of deposited
titanium, the percentage of the covered surface is proportional to the
total amount of deposited titanium, while for higher coverage, the islands grow
stronger in height and therefore the surface coverage increases more
slowly with respect to the volume of titanium, until it finally reaches
100\,\% at around 6.5\,ML. The growth of titanium islands on graphene
will be described elsewhere in more detail.\cite{Mashoff}

In order to investigate hydrogen storage in the Ti--covered samples,
they were exposed to molecular deuterium ($D_2$) for five minutes at a
pressure of $3.5\times 10^{-8}\,mbar$ at room temperature, and then
positioned in front of the mass spectrometer. When the pressure was down
to $2\times 10^{-10}\,mbar$ again, we heated the sample at a constant
rate of $10\,K/s$ up to a temperature of approximately $850\,K$. We
verified by STM that heating the samples up to this temperature does not
significantly change shape or distribution of the titanium islands.

\begin{figure}
  \centerline{\includegraphics[width=0.5\columnwidth]{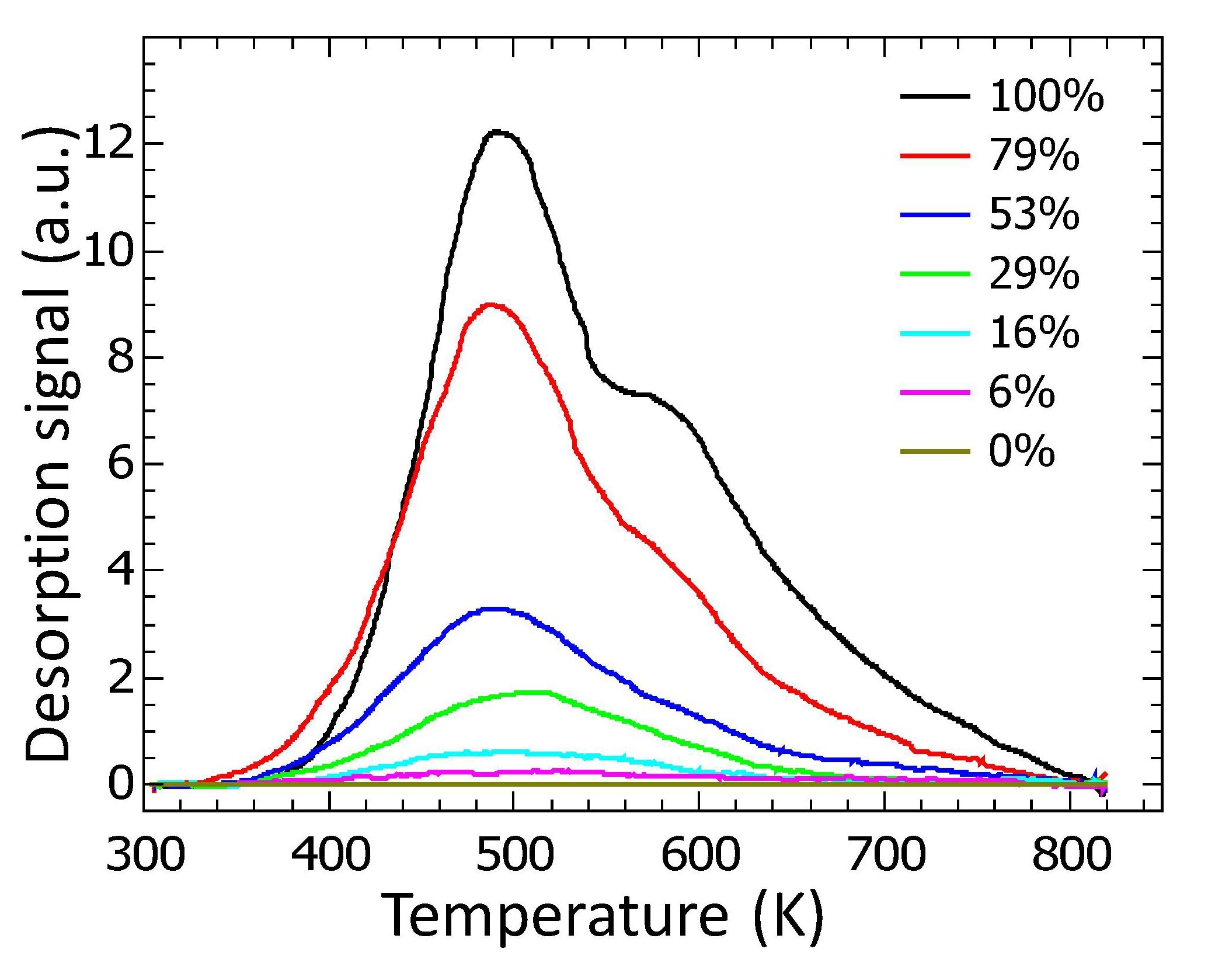}}
  \caption{(Color online) Temperature--dependent desorption spectra of
molecular deuterium (m/e = 4) for titanium  coverages between 0\,\% and
100\,\%. The amount of desorbed deuterium increases with Ti coverage. The desorption starts at around 400\,K and reaches a
maximum around 485\,K. At higher Ti--coverage a shoulder around 560\,K
emerges.}
  \label{curves}
\end{figure}

The resulting temperature--dependent desorption curves of $D_2$ are
shown in Fig.~\ref{curves}. After each measurement, a second ramp cycle
without deuterium was made to verify that all $D_2$ molecules were
actually desorbed during the first cycle. The resulting curves never
showed any signature of $D_2$ desorption. In a further control
experiment, we exposed pure graphene without titanium to a deuterium
flux and did not see any indication of $D_2$ storage (see the 0\,\%
coverage curve in Fig.~\ref{curves}). This is consistent with the
well--known fact that although graphene is capable of binding atomic
hydrogen,\cite{Goler:JPCC} molecular hydrogen does not stick to a
graphene surface at room temperature.\cite{Tozzini} At low titanium
coverage, only a small desorption peak can be detected, but with
increasing Ti coverage, peak amplitude increases, and a shoulder appears
at higher temperatures. The position of these features appears to be
independent of titanium coverage. A simple numerical analysis (a fit
with two Gaussian curves) leads to temperatures of around $(485\pm
5)$\,K for the main peak and around $(560\pm 10)$\,K for the shoulder.

A detailed quantitative analysis of these spectra goes beyond the scope
of this letter; however, even from a qualitative standpoint, very useful
information can be extracted. First, Ti islands on graphene are indeed
capable of storing hydrogen, and the stored amount, which is
proportional to the area under the desorption curve, increases with the
amount of deposited titanium. Second, the measured desorption temperatures
indicate that hydrogen binding is stable at room temperature
and that it desorbs at moderate temperatures close to the target temperature for
practical applications. Third, the desorption energy barrier can be
estimated using the Redhead equation (assuming first order kinetics):\cite{Woodruff}
$E_d/RT=ln(10^{13}/\beta\times T)- 3.64$ with $\beta=dT/dt$ the heating
rate (here 10\,K/s), $E_d$ denotes the desorption energy barrier, $R$
the gas constant, and $T$ the measured TDS peak temperatures. This leads
to values for $E_d$ in the range of 1--1.5\,eV.

Further insight into the behavior of hydrogen with titanium on graphene
can be obtained from a comparison with data available in the literature.
Titanium is well known for its catalytic effect in dissociating hydrogen
molecules into atoms,\cite{JNT831872,PhysRevB.72.153403} similar to what
was reported for Pd.\cite{Cabria2012} These hydrogen atoms can be
expected to form bonds either to titanium or to graphene. In the case of
binding to graphene, the desorption peak would be expected to decrease
when titanium coverage approaches 100\,\%, since there is less and less
uncovered graphene surface available. This is not consistent with our
data. This suggests that hydrogen is actually binding to
titanium. The latter statement is supported by the desorption energy
value estimated from our experiments. The binding energy for a dissociated
hydrogen molecule on a titanium atom on a $C_{60}$-molecule was
calculated to be 1.16\,eV,\cite{PhysRevB.72.153403} which corresponds to
approximately 450\,K using the Redhead equation. Even though the system
is not identical to ours, we expect that binding energies and desorption
temperatures be in the same range, as is actually the case.
Alternatively hydrogen may also be stored in the titanium islands and
forming a titanium hydride complex.\cite{Dhilip2008} The effusion of
deuterium from titanium surfaces was studied before by TDS. The
desorption spectra of deuterated titanium thin films show a desorption
peak around 550\,K,\cite{PhysRevB.58.4130,Nowicka1520} again
consistently with our data.

In summary, the present measurements demonstrate the suitability of
titanium--covered graphene surfaces as a system for hydrogen storage.
Clear evidence of stored hydrogen in our samples was reported, based on TDS results. The
measured desorption temperatures are high enough for stable hydrogen
binding at room temperature and low enough to recover the hydrogen at moderate
temperatures which render titanium--functionalized graphene a very interesting material system for practical applications in hydrogen storage.

\begin{acknowledgments}
We gratefully acknowledge helpful discussions with Erik Vesselli, Georg Held, Vittorio Pellegrini, and Valentina Tozzini.
\end{acknowledgments}

\bibliography{biblio}

\newpage

\end{document}